\documentclass[prd,showpacs,preprint,groupaddress]{revtex4-1}
\usepackage{graphicx}
\usepackage{latexsym}
\usepackage{amsmath}
\usepackage{subfigure}

\begin{document}

\title{The limit on $w'$ for tachyon dark energy}
%\begin{CJK}{GB}{gbsn}
\author{Ximing Chen}
 \email{chenxm@cqupt.edu.cn}
 \affiliation{College of Mathematics and Physics, Chongqing University of Posts and Telecommunications,
Chongqing 400065, China}
\affiliation{State Key Laboratory of Theoretical Physics, Institute of Theoretical Physics, Chinese Academy of Sciences, Beijing 100190, China}
\author{Yungui Gong}
\email{yggong@mail.hust.edu.cn}
 \affiliation{MOE Key Laboratory of Fundamental Quantities Measurement, School of Physics,
 Huazhong University of Science and Technology, Wuhan, Hubei 430074, China}
\affiliation{State Key Laboratory of Theoretical Physics, Institute of Theoretical Physics, Chinese Academy of Sciences, Beijing 100190, China}

\begin{abstract}
We get the same degeneracy relation between $w_0$ and $w_a$ for the tachyon fields as for quintessence and phantom fields.
Our results show that the dynamics of scalar fields with different origins becomes indistinguishable
when the equation of state parameter $w$ does not deviate too
far away from -1, and the time variation $w'$ 
satisfies the same bound for the same class of models.
For the tachyon fields, a limit on $w'$ exists due to the Hubble damping and
we derived the generic bounds on $w'$ for different classes of models.
We may distinguish different models in the phase plane of $w-w'$. The current constraints on $w$ and $w'$
are consistent with all classes of models. We need to improve the constraint on $w'$ by 50\% to distinguish different models.
\end{abstract}


%\preprint{arXiv: }

\maketitle
%\end{CJK}

\section{Introduction}
To explain the cosmic acceleration found by the observations of
type Ia supernovae (SNe Ia) in 1998 \cite{acc0,hzsst98,scpsn98},
we usually introduce dark energy with negative pressure which
consists about 70\% of the total energy density in the universe.
The cosmological constant with the constant equation of state
$w=-1$ is the simplest candidate of dark energy and it is consistent
with current observational data, but its value is too small compared with that estimated from
the vacuum energy of quantum field theory. Dynamical fields with scalar field such as
quintessence \cite{peebles88,wetterich88,quintessence,track1,track2}, phantom \cite{phantom},
tachyon \cite{Sen:2002in,Sen:2002nu,Padmanabhan:2002cp} and k-essence \cite{ArmendarizPicon:2000dh}  were proposed for dark energy.

The scalar field rolls down a very shallow potential, so its equation of state $w$ will approach $-1$ and it dominates
the universe only recently. According to its evolution, the scalar field was classified to several categories.
If the scalar field stuck in a local minimum and starts to roll down to the true minimum until recently, the dark energy models
are called thawing models \cite{Caldwell:2005tm}. For thawing models, $w$ starts with the value -1 at early times and then deviates from -1
at the present time. If the scalar field already rolls down its potential minimum before the onset of acceleration
and it slows down as it comes to dominate the universe, the models are called freezing models \cite{Caldwell:2005tm}.
For the freezing models, $w$ is different from -1 at early times and approaches -1 at the present time.
If $w>-1$ initially and $w$ decreases as the scalar field rolls down its potential, 
the models are called cooling models \cite{Barger:2005sb}.
For the tracking models, the equation of state $w$ tracks below the background equation of state
for a wide range of initial conditions, the tracker fields become dominate only recently
and then it slows to crawl with $w\rightarrow -1$ as $\Omega_\phi\rightarrow 1$ asymptotically.
Both the freezing and tracking models are special cases of cooling models. Due to the Hubble damping and the shallow potential,
the variation of $w$ is bounded \cite{Caldwell:2005tm,Barger:2005sb,Scherrer:2005je,chiba06,chiba06a,Ali:2009mr,Cahn:2008gk,Lopez:2011ur}.
For the thawing quintessence models, the bound was derived as $1+w<w'<3(1+w)$ \cite{Caldwell:2005tm},
where $w'=dw/d\ln a$. For the freezing quintessence models, 
the bound is given as $3w(1+w)<w'\lesssim 0.2w(1+w)$ \cite{Caldwell:2005tm}.
For the tracker quintessence models, a lower bound $-(1-w)(1+w)<w'$ was given in \cite{Scherrer:2005je}. 
Chiba derived a tighter bound $w'>3w(1-w^2)/(1-2w)$ for the tracker quintessence, 
$w'<3w(1-w^2)/(1-2w)$ for the tracker phantom
and $w'>3w(c^2_s-w)(1+w)/(1-2w)$ for the tracker k-essence  \cite{chiba06,chiba06a}.
For the tachyon field, the bound was derived as $-0.8(1+w)<w'<3(1+w)$ \cite{Ali:2009mr}.

Since the scalar field does not change much for most cases, the dynamics of the potential
may be approximated by the linear expansion of $\kappa(\phi)=-3H\dot\phi/V(\phi)$
as $\kappa(\phi)=\kappa_0+\kappa_1(\phi-\phi_0)$ \cite{Crittenden:2007yy}.
Furthermore, for nearly flat potentials,
a general approximate relation between $w$ and $\Omega_\phi$ was found  \cite{Ali:2009mr,Robert2008,Robert,Sourish,delCampo:2010fg}.
The existence of a relation between $w$ and $\Omega_\phi$ was first pointed out for tracker field models in \cite{track1}.
Because of the relation, if we take the zeroth order approximation for $\Omega_\phi$, then we get approximate $w(a)$.
Efstathiou approximated $w(z)$ as $w(z)=w_0-\alpha\ln(1+z)$ in the redshift range $z\lesssim 4$ \cite{efstathiou},
and the approximation was later generalized as the so called Chevallier-Polarski-Linder (CPL) parametrization
with $w(a)=w_0+w_a(1-a)$ \citep{cpl1,cpl2}. Applying the generic relation between $w$ and $\Omega_\phi$ to
the CPL parametrization, analytical degenerated relations between $w_0$ and $w_a$ were derived in \cite{gong1212,Gong:2013bn}
and the two-parameter model reduced to one-parameter SSLCPL model.
In this paper, we discuss the approximate relation between $w$ and $\Omega_\phi$ and the limits on
$w'$ for tachyon fields. In section II, we review the dynamics of tachyon fields. General bounds along
with the bounds for both the thawing and freezing models were derived in section III. In section IV,
we derive the tighter bounds for tracker fields. Conclusions are drawn in section V.

\section{Dynamics of tachyon dark energy}
We start with the Dirac-Born-Infeld type of action for the tachyon field \cite{Sen:2002in,Sen:2002nu}
\begin{equation}
\label{action}
S=\int{-V(\phi)\sqrt{1-\epsilon\partial^{\mu}\phi\partial_{\mu}\phi}\sqrt{-g}d^{4}x},
\end{equation}
where $\epsilon=\pm 1$, and the negative sign corresponds to the phantom type tachyon fields phenomenologically.
If we take the Friedman-Robertson-Walker metric, then the pressure and energy density of the tachyon field $\phi$ are given by
\begin{gather}
\label{tachyonp}
p_t=\frac{V(\phi)}{\sqrt{1-\epsilon\dot{\phi}^2}},\\
\label{tachyonrho}
  \rho_t= -V(\phi)\sqrt{1-\epsilon\dot{\phi}^2}.
\end{gather}
The equation of state for the tachyon field is
\begin{equation}
\label{tachyonw}
w=\frac{p_t}{\rho_t}=\epsilon\dot\phi^2-1.
\end{equation}
So $w<-1$ when $\epsilon=-1$ and $w>-1$ when $\epsilon=1$. For convenience,
we introduce the parameter $\gamma_t=1+w=\epsilon\dot\phi^2$ for the tachyon field $\phi$.
The equation of motion for the tachyon field is
\begin{equation}
\label{sclrw}
\ddot{\phi}+3H\dot{\phi}(1-\epsilon\dot{\phi}^2)+\epsilon\frac{{V^{\prime}}}{V}(1-\epsilon\dot{\phi}^2)=0,
\end{equation}
where $V'=dV/d\phi$. The Hubble parameter $H$ satisfies the Friedmann equations:
\begin{equation}
\label{FR1}
H^{2}=\frac{1}{3}(\rho_{B}+\rho_{\phi}),
\end{equation}
\begin{equation}
\label{hubeq}
\frac{\dot{H}}{H^2}=\frac{3}{2} [\Omega_{\phi}(\gamma_B-\gamma_t)-\gamma_B],
\end{equation}
where $\rho_B$ is the energy density of matter with constant equation of state $w_B$, $\gamma_B=1+w_B$,
$\Omega_{\phi}=\rho_{\phi}/(3H^2)$, $\Omega_{B}=\rho_{B}/(3H^2)$, and we use the unit $8\pi G=1$. For dust,
$w_B=0$ and $\gamma_B=1$.

By using the following dimensionless variables:
\begin{equation}
\label{dim1}
x=\dot{\phi},\quad y=\frac{\sqrt{V(\phi)}}{\sqrt{3}H}, \quad \lambda_t=-\frac{1}{V^{3/2}}\frac{dV}{d\phi},\quad
\Gamma=V\frac{d^2V}{d\phi^2}/\left(\frac{dV}{d\phi}\right)^2,
\end{equation}
the evolution equations can be written as \cite{Sen:2009yh,Ali:2009mr}
\begin{equation}
\label{auto1}
x^{\prime}=-(1-\epsilon x^2)(3x-\sqrt{3}\epsilon\lambda_t y),
\end{equation}
\begin{equation}\label{auto2}
y^{\prime}=\frac{y}{2}[-\sqrt{3}\lambda_t xy-\frac{3(\gamma_{B}-\epsilon x^2)y^2}{\sqrt{1-\epsilon x^2}}+3\gamma_{B}],
\end{equation}
\begin{equation}\label{auto3}
\lambda^{\prime}_t=-\sqrt{3}\lambda^2_t xy(\Gamma-\frac{3}{2}),
\end{equation}
where prime denotes the derivative with respect to $\ln a$.
In terms of the variables $x$ and $y$, we have
\begin{equation}
\label{physvar}
\Omega_{\phi}=\frac{y^2}{\sqrt{1-\epsilon x^2}}, \quad \gamma_t= \epsilon \dot{\phi}^2= \epsilon x^2.
\end{equation}
With the fractional energy density $\Omega_\phi$ and the equation of state parameter $\gamma_t$,
the autonomous Eqs. (\ref{auto1}), (\ref{auto2}) and (\ref{auto3}) can be expressed as
\begin{equation}
\label{auto4}
\gamma_{t}^{\prime}=-6\gamma_{t}(1- \gamma_{t})\pm 2\lambda_t \sqrt{3\epsilon \gamma_{t}\Omega_{\phi}}(1-\gamma_{t})^{5/4},
\end{equation}
\begin{equation}
\label{auto5}
\Omega_{\phi}^{\prime}=3(\gamma_{B}- \gamma_{t})\Omega_{\phi}(1-\Omega_{\phi}),
\end{equation}
\begin{equation}
\label{auto6}
\lambda^{\prime}_t=-\sqrt{3\epsilon \gamma_{t}\Omega_{\phi}}\lambda^2_t (1-\gamma_{t})^{1/4}(\Gamma-\frac{3}{2}),
\end{equation}
where the positive sign corresponds to $\dot\phi>0$ and the negative sign corresponds to $\dot\phi<0$.
As will be discussed below, we can always take the positive sign without loss of generality.
Combining the first two Eqs. (\ref{auto4}) and (\ref{auto5}), we get
\begin{equation}
\label{dgamadomga}
\frac{d\gamma_t}{d\Omega_\phi}=\frac{-6\gamma_{t}(1- \gamma_{t})+2\sqrt{3\epsilon \gamma_{t}\Omega_{\phi}}\lambda_t (1-\gamma_{t})^{5/4}}{3(\gamma_{B}- \gamma_{t})\Omega_{\phi}(1-\Omega_{\phi})}.
\end{equation}
Since we are interested in the dynamics which is close to a cosmological constant, so $\gamma_t\ll 1$ and the potential changes very slowly
with a nearly constant slope $\lambda_t\approx \lambda_0$. With these approximations, and take dust $\gamma_B=1$, we get
\begin{equation}
\label{dgamadomga1}
\frac{d\gamma_t}{d\Omega_\phi}=\frac{-6\gamma_t+2\lambda_0\sqrt{3\epsilon\gamma_t\Omega_\phi}}{3\Omega_\phi(1-\Omega_\phi)}.
\end{equation}
The above equation obtained for the tachyon fields is the same as that obtained for the quintessence and phantom fields. This shows that
the dynamics is quite generic for general scalar fields, so it is hard to distinguish different dynamics that caused the cosmic acceleration.
The solution to the above Eq. (\ref{dgamadomga1}) for general thawing models is \cite{Robert2008,Robert,Ali:2009mr}
\begin{equation}
\label{dynasol1}
\gamma_t=\frac{1}{3}\epsilon\lambda_0^2\left[\frac{1}{\sqrt{\Omega_\phi}}-\left(\frac{1}{\Omega_\phi}-1\right)\tanh^{-1}(\sqrt{\Omega_\phi})\right]^2,
\end{equation}
and
\begin{equation}
\label{dynasol2}
\lambda_0^2=3\epsilon\gamma_{t0}\left[\frac{1}{\sqrt{\Omega_{\phi0}}}-(\Omega_{\phi0}^{-1}-1)\tanh^{-1}
\sqrt{\Omega_{\phi0}}\right]^{-2}.
\end{equation}
Substituting the solution to Eq. (\ref{auto5}) with the approximation $\gamma_t\ll 1$,
\begin{equation}
\label{wapprox1}
\Omega_\phi(a)\approx [1+(\Omega_{\phi 0}^{-1}-1)a^{-3}]^{-1},
\end{equation}
into Eq. (\ref{dynasol1}), we get the evolution of the equation of state parameter $\gamma_t(a)$ for the tachyon dark energy.
If we take further approximation,
i.e., we take Taylor expansion of $\Omega_\phi$ around $a=1$ from Eq. (\ref{wapprox1}), we get
\begin{equation}
\label{omegaphieq1}
\Omega_\phi(a)\approx \Omega_{\phi 0}+3\Omega_{\phi 0}(1-\Omega_{\phi 0})w_{0} (1-a).
\end{equation}
Using Eqs. (\ref{dynasol1}), (\ref{dynasol2}) and (\ref{omegaphieq1}),
we get the SSLCPL parametrization $w(a)=w_0+w_a(1-a)$ with \cite{gong1212,Gong:2013bn}
\begin{equation}
\label{waapproxeq1}
w_a=-6w_0(1+w_0)\frac{(\Omega_{\phi 0}^{-1}-1)[\sqrt{\Omega_{\phi0}}-\tanh^{-1}(\sqrt{\Omega_{\phi0}})]}
{\Omega_{\phi 0}^{-1/2}-(\Omega_{\phi 0}^{-1}-1)\tanh^{-1}(\sqrt{\Omega_{\phi0}})}.
\end{equation}
We see that the degeneracy relation (\ref{waapproxeq1}) among $w_0$, $w_a$ and $\Omega_{\phi 0}$
is quite general, it holds for both canonical and tachyon scalar fields with $w\le -1$ and $w\ge -1$.

\section{General bound on $w'$}

For $w\ge -1$, the scalar field rolls down the potential, $dV/dt=(dV/d\phi) \dot\phi<0$, so $\dot\phi$ and $\lambda_t$ have
the same sign. Therefore, the second term on the right hand side of Eq. (\ref{auto4})
is always positive, and we can take the positive sign only. Therefore, $\gamma_t'> -6\gamma_t(1-\gamma_t)$.
For $w\ge -1$, one has a lower bound on $w'$ as
\begin{equation}
\label{w-exprr}
w'>6w(1+w).
\end{equation}
For the phantom case $w<-1$, $dV/dt=(dV/d\phi) \dot\phi>0$, so $\dot\phi$ and $\lambda_t$ have
the opposite sign, and the second term on the right hand side of Eq. (\ref{auto4})
is always negative. Therefore, for $w<-1$, one gets an upper bound
\begin{equation}
\label{phantbld1}
w'<6w(1+w).
\end{equation}

On the other hand, the equation of motion for the tachyon field gives
\begin{equation}
\label{tachwdereq}
w'=2(1+w)\frac{\ddot\phi}{H\dot\phi}.
\end{equation}
The scalar field does not change much during the matter domination
and starts to change when it becomes dominant, so the acceleration of the scalar
field is limited by the Hubble damping as
\begin{equation}
\label{phidereq3}
|\ddot\phi|\lesssim \frac{|\dot\phi|}{t}=\frac{3}{2}H|\dot\phi|.
\end{equation}
For the thawing model, we have $\ddot\phi<3H\dot\phi/2$. For the freezing model, we have $\ddot\phi>-3H\dot\phi/2$.
Therefore, we have more stringent bound for some special cases.
For the thawing model with $w>-1$, we have
\begin{equation}
\label{wblda}
3(1+w)>w'>0.
\end{equation}
In \cite{Ali:2009mr}, the authors derived the bound $-0.8(1+w)<w'<3(1+w)$ for thawing tachyon field. However,
for thawing field, $w'>0$, so the lower bound should be $w'>0$.
The thawing solution (\ref{waapproxeq1}) for $w_a$ gives $3w_0(1+w_0)<w_a<0$. Since $w'=-w_a$, so the bound on $w'$ becomes
$0<w'<-3w_0(1+w_0)<3(1+w_0)$ which is consistent with the bound (\ref{wblda}).

For the thawing model with $w<-1$, we have
\begin{equation}
\label{wbldb}
0>w'>3(1+w).
\end{equation}
The thawing solution (\ref{waapproxeq1}) gives $3w_0(1+w_0)>w_a>0$ which leads to the bound
$0>w'>-3w_0(1+w_0)>3(1+w_0)$. Therefore the thawing solution (\ref{waapproxeq1}) is consistent with the bound (\ref{wbldb}).

For the freezing model with $w>-1$, we have
\begin{equation}
\label{wbldc}
-3(1+w)<w'<0.
\end{equation}
Since we are interested in the region when $w$ is not far away from -1, $w<-1/2$,
so $-3(1+w)>6w(1+w)$ and this lower bound $-3(1+w)$ is more restrictive compared with the general bound (\ref{w-exprr}).
For some cases, the model may stuck in the point with $w=0$ initially, then $w$ decreases to $-1$ very fast
and the lower bound (\ref{wbldc}) can be violated, but the general bound (\ref{w-exprr}) is still satisfied.

For the freezing model with $w<-1$, we have
\begin{equation}
\label{wbldd}
-3(1+w)>w'>0.
\end{equation}
Because $-3(1+w)<6w(1+w)$, 
so this upper bound $-3(1+w)$ is more restrictive compared with the general bound (\ref{phantbld1}).

\section{Bounds on $w'$ for tracker fields}

Similar to the quintessence and phantom cases, we derive the ``tracker equation" \cite{track2,Rubano:2003et} for the tachyon field first.
Combining Eqs. (\ref{auto4}), (\ref{auto5}) and (\ref{auto6}), we obtain
\begin{equation}
\label{tracker}
\Gamma-\frac{3}{2}=-\frac{(1-\gamma_{t})[2\gamma''_t+3(2-\gamma_{t})\gamma_t']}{[\gamma_{t} '+6\gamma_{t}(1-\gamma_{t})]^2}
+\frac{(1-\frac{7}{2}\gamma_{t})\gamma_{t}^{\prime 2}}{\gamma_{t}[\gamma_{t} '+6\gamma_{t}(1-\gamma_{t})]^2}
+\frac{3(1-\gamma_{t})(\gamma_B-\gamma_{t})(1-\Omega_{\phi})}{\gamma_{t}'+6\gamma_{t}(1-\gamma_{t})}.
\end{equation}
The ``tracker equation" (\ref{tracker}) is independent of $\epsilon$, so it holds for both the phantom and non-phantom cases.
To simplify the above ``tracker equation", we introduce the variable $X$ as
\begin{equation}
\label{Xvareq}
X=\ln{\left[\frac{\gamma_{t}}{\epsilon(1-\gamma_{t})}\right]}=\ln{\left(\frac{1+w}{-\epsilon w}\right)}.
\end{equation}
So
\begin{equation}
\label{gprime}
X'=\frac{\gamma_{t}'}{\gamma_{t}(1-\gamma_{t})}=-\frac{w'}{w(1+w)}.
\end{equation}
Using the variable $X$, the ``tracker equation"  (\ref{tracker}) becomes
\begin{equation}
\label{tracker1}
\Gamma-\frac{3}{2}=-\frac{2X''}{(1+w)(6+X')^2}-\frac{(1-w)X'}
{2(1+w)(6+X')}+\frac{3(w_B-w)(1-\Omega_{\phi})}{(1+w)(6+X')}.
\end{equation}
For the tracker solution, initially $\Omega_{\phi}$ is negligible and $w$ is almost a constant. As the scalar field
becomes dominant, $w$ decreases and approaches toward -1 asymptotically,
so it belongs to the class of freezing models. However, because of the $w-\Omega_\phi$ relation \cite{track1} 
for the tracker solution and $\Omega_{\phi 0}\sim 0.7$, $w_0$ is different from $-1$ which makes 
the tracking model distinguishable from $\Lambda$CDM model.
By taking $X''=X'=0$ and $\Omega_{\phi}=0$ in Eq. (\ref{tracker1}), we get
 \begin{equation}
 \label{w-express}
w=\frac{w_B-2(\Gamma-\frac{3}{2})}{1+2(\Gamma-\frac{3}{2})}.
\end{equation}
For the dust $w_B=0$, we get the track solution
\begin{equation}
\label{w-express1}
w_{trk}=-2\frac{(\Gamma-\frac{3}{2})}{1+2(\Gamma-\frac{3}{2})}.
\end{equation}
So the nearly constant $\Gamma=3/2-w_{trk}/2(1+w_{trk})$, $\Gamma>3/2$ for $w>-1$ and $\Gamma<1$ for $w<-1$.
Since $X'$ will stop decreasing and then increases toward zero, it has a minimum.
To find the minimum, we set $X''=0$ in Eq. (\ref{tracker1}), then we get
\begin{equation}
\label{X-minimum}
X'_{m}=-6\frac{w(1-\Omega_{\phi})+2(1+w)(\Gamma-\frac{3}{2})}{(1-w)+2(1+w)(\Gamma-\frac{3}{2})}.
\end{equation}
Note that $\Omega_{\phi}<1$ and $w<0$, a lower bound on $X_{m}'$ is 
\begin{equation}
\label{X-minimum1}
X'_{m}>-\frac{12(1+w)(\Gamma-\frac{3}{2})}{(1-w)+2(1+w)(\Gamma-\frac{3}{2})}.
\end{equation}
Because the right hand side of Eq. (\ref{X-minimum1}) is an decreasing function of $w$, and $w$
decreases from the tracker solution (\ref{w-express1}), so we get a lower bound on $X'_{m}$ by replacing $w$ with $w_{trk}$,
\begin{equation}
\label{X-minimum2}
X'_{m}>\frac{6w_{trk}}{1-2w_{trk}}>\frac{6w}{1-2w}.
\end{equation}
In the last inequality, we notice that the function $6w/(1-2w)$ decreases as $w$ decreases.
For the case of $w>-1$, we get a lower bound on $w'$,
\begin{equation}
\label{w-limit}
w'>-\frac{6w^2}{1-2w}(1+w)
=\frac{3w}{1-2w}(1+w)(c_s^2-w),
\end{equation}
where the sound speed of the tachyon field $c_s^2=-w$.
The bound was also derived for general tracker K-essence with $w>-1$ in \cite{chiba06}.
Since $3w/(1-2w)>-1$, so the lower bound can be also written as
\begin{equation}
\label{w-limit1}
w'>2w(1+w).
\end{equation}
The lower bound $2w(1+w)>-3(1+w)$ is more stringent than the bound (\ref{wbldc}) for general freezing model with $w>-1$.

\begin{figure}[htp]
\centerline{\includegraphics[width=0.5\textwidth]{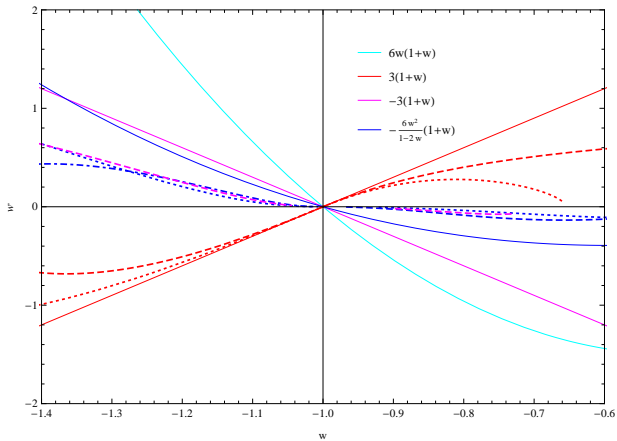}}
\caption{The bounds on $w'$  for different models. 
We also plot the trajectories of different models with the potential $V(\phi)=\phi^n$ by numerically solving Eqs. (\ref{auto4})-(\ref{auto6}),
here $n$ can be either positive or negative. The cyan line corresponds to the general bound, the red lines correspond to the thawing
models, the magenta lines correspond to the freezing models, and the blue lines correspond to the tracking models.}
\label{bounds}
\end{figure}

For $w<-1$, we get an upper bound
\begin{equation}
\label{w-expq}
w'<-\frac{6w^2}{1-2w}(1+w).
\end{equation}
Again for the case that $w$ is not far away from -1, this upper bound is 
more restrictive than the bound (\ref{wbldd}) for general freezing model with $w<-1$.
The general bounds (\ref{w-exprr}) and (\ref{phantbld1}), the thawing bounds (\ref{wblda}) and (\ref{wbldb}), 
the freezing bounds (\ref{wbldc}) and (\ref{wbldd}), and the tracker bounds (\ref{w-limit}) and (\ref{w-expq}) are shown 
in Fig. \ref{bounds}. We also numerically solve Eqs. (\ref{auto4})-(\ref{auto6}) for the potential $V(\phi)=\phi^n$
to show explicitly that those bounds are satisfied.

\begin{figure}[htp]
\centerline{\includegraphics[width=0.6\textwidth]{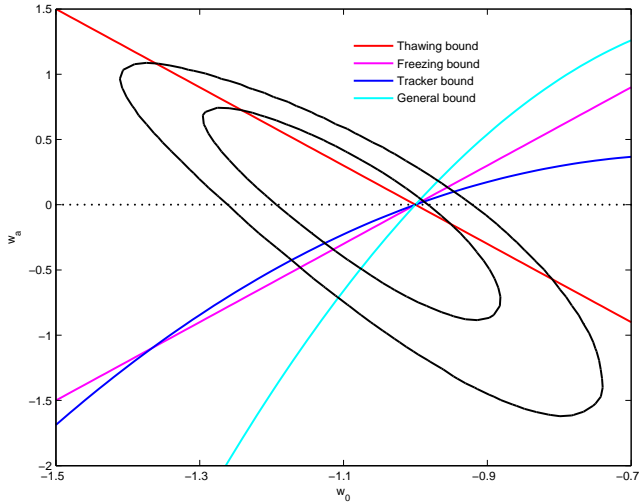}}
\caption{The marginalized $1\sigma$ and $2\sigma$ contours of $w_0$ and $w_a$ of the flat CPL model from observational data \cite{Gao:2013pfa}.
We also show the bounds we derived.}
\label{w0acont}
\end{figure}

To see whether it is possible to distinguish different dynamical models from the above bounds we derived, we use the observational constraints on the
flat CPL model \cite{Gao:2013pfa}. For CPL parametrization, $w'(a=1)=-w_a$. In Fig. \ref{w0acont}, we show the marginalized $1\sigma$ and $2\sigma$ contours of $w_0$ and $w_a$ from the combined SNLS3 SNe Ia, {\em Planck}, BAO and $H(z)$ data \cite{Gao:2013pfa}.
We also plot the bounds (\ref{w-exprr}),
(\ref{phantbld1}), (\ref{wblda}), (\ref{wbldb}), (\ref{wbldc}), (\ref{wbldd}), (\ref{w-limit}) and (\ref{w-expq}) in Fig. \ref{w0acont}.
Most of the contours are inside the general bounds (\ref{w-exprr}) and (\ref{phantbld1}), but only a small part of the contours
satisfy the bounds  (\ref{wblda}), (\ref{wbldc}) and (\ref{w-limit}) for $w>-1$. The observational constraints
slightly favor $w_a<0$ or $w'>0$ when $w>-1$, so the thawing model is slightly favored.

\section{Conclusions}

The generic relation (\ref{dynasol1}) between $w$ and $\Omega_\phi$ holds for the qintessence,
phantom and tachyon thawing fields. Based on the generic relation, we derived the SSLCPL model
with analytic relation (\ref{waapproxeq1}) for $w_0$ and $w_a$ in the context of tachyon field.
The same SSLCPL parametrization holds for thawing quintessence, phantom, tachyons
with $w>-1$ and $w<-1$ no matter what the form the potential takes.
This shows that the dynamics of scalar fields become indistinguishable when $w$ does not deviate too
far away from -1, and makes the models less distinguishable from $\Lambda$CDM model. 
For the tachyon fields, we also derived the general bounds (\ref{w-exprr}) and
(\ref{phantbld1}), the bounds (\ref{wblda}) and (\ref{wbldb}) for the thawing models,
and the bounds (\ref{wbldc}) and (\ref{wbldd}) for the freezing models.
Starting from the ``tracker equation" (\ref{tracker1}) we derived for tachyon fields,
tighter bounds (\ref{w-limit}) and (\ref{w-expq}) were given for the tracking models.
The typical bound on $w'$ is $\sim w(1+w)$ and the tightest bound is around $2w(1+w)$ for the tracker fields.
If $w$ is measured to be $-1$ within 20\% error, then we require $\delta w'\sim 0.4$ to detect tracking models.
The current observational constraints on the flat CPL model are
$w_0=-1.10^{+0.17}_{-0.10}$ and $w_a=0.08^{+0.32}_{-0.76}$ \cite{Gao:2013pfa}.
so $w_0$ is constrained to be $w_0>-1$ by 10\% and $w_0<-1$ by 20\%,
but the uncertainties on $w_a$ are $\delta w_a\sim 0.4$ and $\delta w_a\sim -0.7$.
Therefore, the current data cannot distinguish different classes of models.
We need the data to constrain $w_a$ with the error bar better
than $|\delta w_a| \sim 0.4$, i.e., we need to improve the constraint on $w_a$ by up to 50\%.

\begin{acknowledgments}
This work was partially supported by the National Basic Science Program (Project 973) of
China under grant No. 2010CB833004, the NNSF project of China under
grant Nos. 10935013 and 11175270,
the Program for New Century Excellent Talents in University,
the Fundamental Research Funds for the Central Universities and CQ CMEC under grant No. KJ110523.
\end{acknowledgments}

%\bibliographystyle{h-physrev}
%\bibliographystyle{physlett}
%\bibliography{../book/cosmologyref}

\begin{thebibliography}{10}

\bibitem{acc0}
S.~Perlmutter {\em et~al.},
 Nature
   391 (1998) 51.

\bibitem{hzsst98}
A.~G. Riess {\em et~al.},
 Astron. J.
   116 (1998) 1009.

\bibitem{scpsn98}
S.~Perlmutter {\em et~al.},
 Astrophys. J.
   517 (1999) 565.

\bibitem{peebles88}
B.~Ratra and P.~Peebles,
 Phys. Rev. D
   37 (1988) 3406.

\bibitem{wetterich88}
C.~Wetterich,
 Nucl. Phys. B
   302 (1988) 668.

\bibitem{quintessence}
R.~Caldwell, R.~Dave, and P.~J. Steinhardt,
 Phys. Rev. Lett.
   80 (1998) 1582.

\bibitem{track1}
I.~Zlatev, L.-M. Wang, and P.~J. Steinhardt,
 Phys. Rev. Lett.
   82 (1999) 896.

\bibitem{track2}
P.~J. Steinhardt, L.-M. Wang, and I.~Zlatev,
 Phys. Rev. D
   59 (1999) 123504.

\bibitem{phantom}
R.~Caldwell,
 Phys. Lett. B
   545 (2002) 23.

\bibitem{Sen:2002in}
A.~Sen,
 JHEP
   0207 (2002) 065.

\bibitem{Sen:2002nu}
A.~Sen,
 JHEP
   0204 (2002) 048.

\bibitem{Padmanabhan:2002cp}
T.~Padmanabhan,
 Phys. Rev. D
   66 (2002) 021301.

\bibitem{ArmendarizPicon:2000dh}
C.~Armendariz-Picon, V.~F. Mukhanov, and P.~J. Steinhardt,
 Phys. Rev. Lett.
   85 (2000) 4438.

\bibitem{Caldwell:2005tm}
R.~Caldwell and E.~V. Linder,
 Phys. Rev. Lett.
   95 (2005) 141301.

\bibitem{Barger:2005sb}
V.~Barger, E.~Guarnaccia, and D.~Marfatia,
 Phys. Lett. B
   635 (2006) 61.

\bibitem{Scherrer:2005je}
R.~J. Scherrer,
 Phys. Rev. D
   73 (2006) 043502.

\bibitem{chiba06}
T.~Chiba,
 Phys. Rev. D
   73 (2006) 063501.

\bibitem{chiba06a}
T.~Chiba,
 Phys. Rev. D
   80 (2009) 129901(E).

\bibitem{Ali:2009mr}
A.~Ali, M.~Sami, and A.~Sen,
 Phys. Rev. D
   79 (2009) 123501.

\bibitem{Cahn:2008gk}
R.~N. Cahn, R.~de~Putter, and E.~V. Linder,
 JCAP
   0811 (2008) 015.

\bibitem{Lopez:2011ur}
L.~A. Urena-Lopez,
 JCAP
   1203 (2012) 035.

\bibitem{Crittenden:2007yy}
R.~Crittenden, E.~Majerotto, and F.~Piazza,
 Phys. Rev. Lett.
   98 (2007) 251301.

\bibitem{Robert2008}
R.~J. Scherrer and A.~Sen,
 Phys. Rev. D.
   77 (2008) 083515.

\bibitem{Robert}
R.~J. Scherrer and A.~Sen,
 Phys. Rev. D.
   78 (2008) 067303.

\bibitem{Sourish}
S.~Dutta and R.~J. Scherrer,
 Phys. Lett. B.
   704 (2011) 265.

\bibitem{delCampo:2010fg}
S.~del Campo, V.~H. Cardenas, and R.~Herrera,
 Phys. Lett. B
   694 (2011) 279.

\bibitem{efstathiou}
G.~Efstathiou,
 Mon. Not. Roy. Astron. Soc.
   310 (1999) 842.

\bibitem{cpl1}
M.~Chevallier and D.~Polarski,
 Int. J. Mod. Phys. D.
   10 (2001) 213.

\bibitem{cpl2}
E.~V. Linder,
 Phys. Rev. Lett.
   90 (2003) 091301.

\bibitem{gong1212}
Q.~Gao and Y.~Gong,
 Int. J. Mod. Phys. D
   22 (2013) 1350035.

\bibitem{Gong:2013bn}
Y.~Gong and Q.~Gao,
 {On the effect of the degeneracy between w0 and wa}
 (2013).

\bibitem{Sen:2009yh}
S.~Sen, A.~Sen, and M.~Sami,
 Phys. Lett. B
   686 (2010) 1.

\bibitem{Rubano:2003et}
C.~Rubano, P.~Scudellaro, E.~Piedipalumbo, S.~Capozziello, and M.~Capone,
 Phys. Rev. D
   69 (2004) 103510.

\bibitem{Gao:2013pfa}
Q.~Gao and Y.~Gong,
 {On the compatibility of different observational data}
 (2013).

\end{thebibliography}

\end{document}